\documentclass[12pt]{article}
\usepackage{graphicx}
\usepackage{caption}
\usepackage[justification=centering]{caption}
\usepackage{multirow}
\usepackage{adjustbox}
\usepackage{array}

\newcommand\pubnumber{WSU-HEP-1507\\DPF2015-233}
\newcommand\pubdate{November 1, 2015}
\usepackage[pdftex]{hyperref}

\def\Title#1{\begin{center} {\Large #1 } \end{center}}

\newcommand\pubblock{\rightline{\begin{tabular}{l} \pubnumber\\
         \pubdate  \end{tabular}}}
\newenvironment{Abstract}{\begin{quotation}  }{\end{quotation}}
\newenvironment{Presented}{\begin{quotation} \begin{center} 
             PRESENTED BY JOYDEEP ROY AT\end{center}\bigskip 
      \begin{center}\begin{large}}{\end{large}\end{center} \end{quotation}}

\begin{document}
\begin{titlepage}
\pubblock

\vfill
\Title{Model Independent Analysis of 
the Proton Magnetic Radius}
\vfill
\begin{center}
{\sc  Joydeep Roy$^{(a)}$, Zachary Epstein$^{(b)}$ }, Gil Paz$^{(a)}$\\
{\it $^{(a)}$
Department of Physics and Astronomy \\
Wayne State University, Detroit, Michigan 48201, USA 
}\\
{\it $^{(b)}$ Department of Physics, University of Maryland, College Park, Maryland 20742, USA
}\\
\end{center}
\vfill
\begin{Abstract}
The proton is a fundamental constituent of matter. It is an extended object with finite size that can be inferred with some degree of accuracy from several measurements. Using constraints from the analytic behavior of the form factors we present here a model-independent study that extracts the proton magnetic radius from scattering data. From electron-proton scattering data we find  $r_M^p = 0.91_{-0.06}^{+0.03} \pm 0.02$ fm. When we include electron-neutron scattering data and $\pi\pi$ data, we find $r_M^p = 0.87_{-0.05}^{+0.04}\pm 0.01$ fm and $r_M^p =0.87_{-0.02}^{+0.02}$ fm respectively. The neutron magnetic radius is extracted as $r_M^n = 0.89_{-0.03}^{+0.03}$ fm combining all three data sets.
\end{Abstract}
\vfill
\begin{Presented}
DPF 2015\\
The Meeting of the American Physical Society\\
Division of Particles and Fields\\
Ann Arbor, Michigan, August 4--8, 2015\\
\end{Presented}
\vfill
\end{titlepage}
\setcounter{footnote}{0}

\section{Introduction}

The proton is a fundamental constituent of matter. It is an extended object with finite size that can be inferred with some degree of accuracy from several measurements. Two ways to describe the proton size are the electric and magnetic radii, see the definitions below.

The electric radius can be extracted from electron-proton scattering experiments,
$(r_E^p = 0.871\pm 0.009$ fm) \cite{Hill:2010yb} and Lamb shift in Muonic Hydrogen $(r_E^p =0.84184\pm 0.0006$ fm) \cite{Pohl:2010zz},$(r_E^p =0.84087\pm 0.00039$ fm) \cite{Antognini:1900ns}.  The reason for the discrepancy between these values, often described as the \textquotedblleft proton radius puzzle", is still unknown. In the literature there also exist several values of the proton magnetic radius. The PDG 2014 \cite{PDG:2014} reports the magnetic radius of the proton as $r_M^p =0.777\pm 0.014$ fm according to the $A1$ collaboration \cite{Bernauer:2010wm}. This value is significantly smaller than the other values such as $r_M^p =0.854\pm 0.005$ fm extracted in \cite{Belushkin:2006qa} or $r_M^p =0.876\pm 0.020$ fm \cite{Borisyuk:2009mg}. We will use the \textquotedblleft $z$-expansion" method originally used in \cite{Hill:2010yb} to extract the magnetic radius of the proton from different scattering data \cite{Epstein:2014zua}. We will also report the neutron magnetic radius extracted using the same technique.

\section{Form factors and Magnetic radius}

Since our analysis is based on the analytic properties of the form factors, it is useful to review some of their features.

\subsection{Definitions of Form factors}
 The proton's extended structure can be probed by an electromagnetic current and is described by two form factors, known as Dirac and Pauli form factors. They are defined by $F_1^N$ and $F_2^N$ respectively \cite{Epstein:2014zua}
\begin{equation}
\langle N(p')|J_\mu^{\rm em}|N(p)\rangle=\bar u(p')
\left[\gamma_\mu F^N_1(q^2)+\frac{i\sigma_{\mu\nu}}{2m_N}F^N_2(q^2)q^\nu\right]u(p)\
\end{equation}
where $q^2=(p'-p)^2=t$ and $N$ stands for $p$ or $n$. The Sachs  electric form factor $(G_E)$ and magnetic form factor $(G_M)$ are related to the Dirac-Pauli basis as \cite{Ernst:1960zza}
\begin{equation}
G^N_E(t) = F^N_1(t) + {t\over 4m_N^2} F^N_2(t) \,, \quad
G^N_M(t) = F^N_1(t) + F^N_2(t) \,. 
\end{equation} 
At $t$ = 0,  $G_E^p(0)$ = 1,  $G_E^n(0)$ = 0,  $G_M^p(0)$ = $\mu_p \approx $ 2.793, $G_M^n(0)$ =  $\mu_n \approx $  -1.913 \cite{Beringer:1900zz}.
\medskip  

The isoscalar and isovector form factors are defined as
\begin{equation} 
  G_{E,M}^{I=0}  =  G_{E,M}^p  + G_{E,M}^{n}\quad ,   \qquad  G_{E,M}^{I=1}  =  G_{E,M}^p - G_{E,M}^n 
\end{equation}
such that at $t$ = 0 they are, $G_M^{I=0}(0) = \mu_p + \mu_n = 0.88$ and 
$G_M^{I=1}(0) = \mu_p - \mu_n = 4.8$. Here $ G_{E,M}^p $ and $ G_{E,M}^n $ stands for the proton and neutron form factors.
The magnetic radius of the proton is defined as $r_M^p \equiv \sqrt{\langle r^2\rangle_M^p}$, where   
\begin{equation}
\langle r^2 \rangle_M^p=
\frac{6}{G_M^p(0)}\frac{d}{dq^2}G_M^p(q^2)\bigg|_{q^2=0}\, .
\end{equation}
\subsection{Analyticity of Form factors}

Since the exact functional behavior of the form factors is not known, the number of the parameters used to fit the  experimental data is not determined a priori. Too many parameters will limit the predictive power and too few will bias the fit. Therefore, our goal is to provide some constraints on the functional behavior of the form factors. For that we use \textquotedblleft $z$-expansion" method \cite{Hill:2010yb}. 

This method is based upon the analytic properties of the form factor $G_M^p$. The form factor is analytic in the complex $t$-plane outside of a cut along the positive $q^2$ axis with a threshold at $4m_{\pi}^2$ and extended upto infinity $( t= 4m_{\pi}^2 , \infty )$. This is shown in Fig.1.

\begin{figure}[h]
  \centering
  \includegraphics [width=120mm]{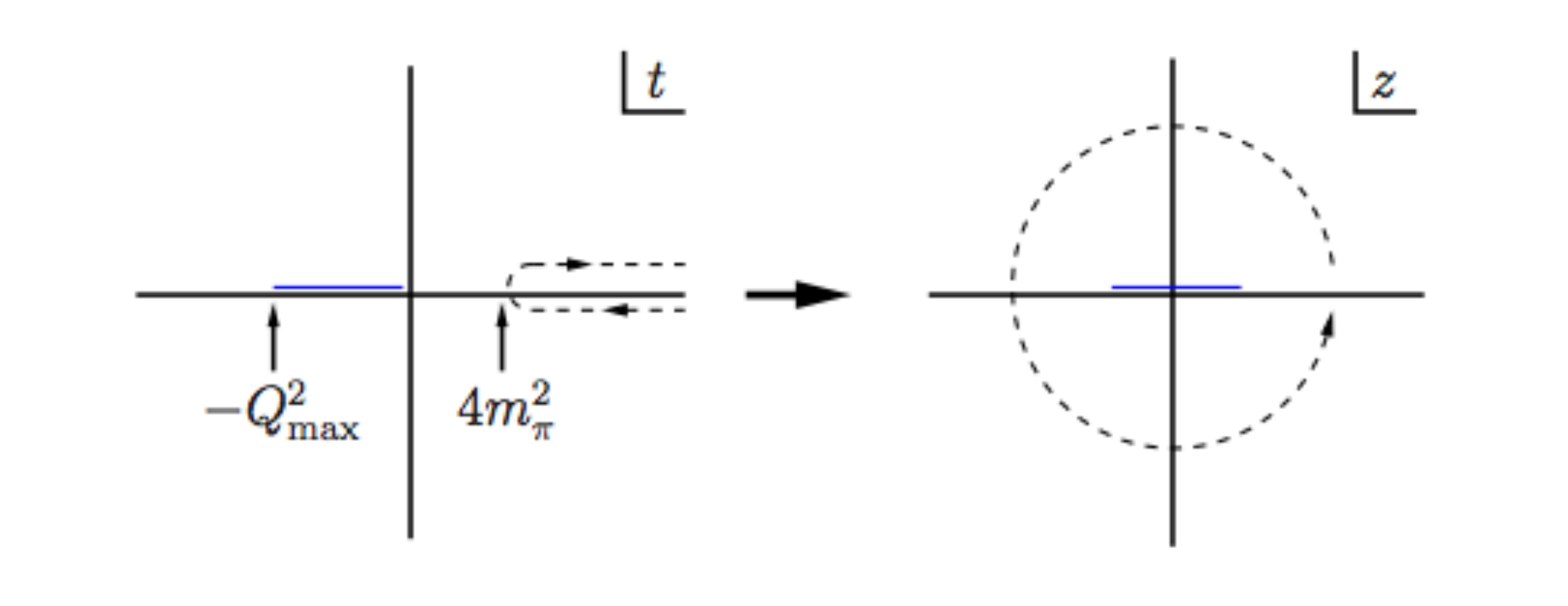}
  \caption{Conformal mapping of the cut plane to the unit circle \cite{Hill:2010yb}}
\end{figure}

So we begin our model-independent approach by performing a conformal mapping of the domain of analyticity onto the unit circle by defining the variable $z$ as 
\begin{equation}
z(t,t_{\rm cut},{t_0}) = \frac{\sqrt{t_{\rm cut}-t} - \sqrt{t_{\rm cut}-t_0}}{\sqrt{t_{\rm cut}-{t}} + \sqrt{t_{\rm cut}-t_0}}
\end{equation}
where for the present case $t_{\rm cut} $= $4m_\pi^2 $  and  ${t_0} $  is a free parameter representing the point mapping onto $z$ = 0 .

Since the values of the form factors at $q^2=0$ are known, it is convenient to use $t_0=0$. The results are independent of this choice \cite{Hill:2010yb}. The form factors can be expanded in a  power series in $z(q^2)$:
\begin{equation}\label{series} 
G(q^2) = \sum_{k=0}^\infty a_k \, z(q^2)^k ,
\end{equation}
where $ z(q^2) = z(q^2,t_{\rm cut},t_0 = 0) $.

\subsection{Coefficients}

The analytic structure in the $t$-plane, illustrated in the Fig.1 implies the dispersion relation,

\begin{equation}
G(t) = \frac{1}{\pi}\int_{t_{\rm cut}}^\infty dt^\prime \frac{\mbox {Im} G(t^\prime + i0)}{t^\prime - t}
\end{equation}
Equation (5) maps points just above (below) the cut in the $t$ plane onto points in the lower (upper) half unit circle in the z plane . Parameterizing the unit circle by $z(t)$ = $e^{i\theta}$ and solving (5) for $t$ with changed limits we find


\begin{eqnarray}
a_0&=&\frac{1}{\pi}\int_0^\pi d\theta\,{\rm Re}\, G[ t(\theta)+i0]=G(t_0) \,,
\nonumber\\
a_k& = & \frac{2}{\pi}\int_{t_{\rm cut}}^{\infty}\frac{dt}{t - t_0}\sqrt{\frac{t_{\rm cut} - t_0}{t - t_{\rm cut}}} \mbox {Im} G[t(\theta)+ i0] \sin((k\theta(t)),\ k \geq1
\end{eqnarray}

Knowledge of the imaginary part of  $G(t)$ helps us put constraints on the coefficients $a_k$.

\subsection{Bounds on the coefficients}

For a realistic extraction of the proton magnetic radius we need to put appropriate bounds on the coefficients $a_k$. In \cite{Hill:2010yb} Hill and Paz showed that in order to extract the electric charge radius $r_E^p$ the bounds $|a_k| \leq 5,10$ are conservative enough.

The vector dominance ansatz \cite{Epstein:2014zua} was used to estimate the size of the $a_k$. Also from eqn.(3) the magnetic form factors at  $q^2=0$ are given by $G_{M}^{(0)}(0)\approx 0.88$ and $G_{M}^{(1)}(0)\approx 4.7$, compared to $G_{E}^{(0,1)}(0)=1$. Since the vector dominance ansatz is normalized by the value at $q^2=0$, coefficients are proportional to this value. Thus we find that $|a_k|\leq1.1$ for $I=0$ and $|a_k|\leq5.1$ for $I=1$. 

Since for the magnetic isovector form factor the singularities that are closest to the cut arise from the two pion continuum we used explicit $\pi\pi$ continuum to find $a_0\approx 7.9$, $a_1\approx -5.5$, $a_2\approx -6.1$, $a_3\approx  -2.9$, $a_4\approx 1.1$ \cite{Epstein:2014zua}. Also for the case of two nucleon threshold we used $e^+e^-\to N\bar N$ data to constrain the magnetic form factor. Calculations using these data showed that the contribution to $|a_k|$ in the region $t\geq 4m_N^2$ can be neglected \cite{Epstein:2014zua}. Therefore we conclude that $|a_k|\leq 5$ is too stringent and we use the looser bounds $|a_k|\leq10$ and $|a_k|\leq15$ as our default.

One can also use a bound on the ratio $|\frac{a_k}{a_0}|$  \cite{Epstein:2014zua}. From the known value of $a_0$ this translates to a bound on $|a_k|$. In \cite{Epstein:2014zua} it was shown that the results from bounds on the ratio are consistent with the values extracted using the default bounds. The same is true for a higher bound like $|a_k|\leq20$.

\section{Proton magnetic radius extraction} 
\subsection{Proton data}
For our data fitting we use (\ref{series}). We fit our data by minimizing a $ \chi^2 $ function \cite{Hill:2010yb} ,
\begin{equation}\label{chi2}
\chi^2=\sum_i(\mbox{data}_{\,i}- \mbox{theory}_i)^2/(\sigma_i)^2,
\end{equation}
Where $i$ ranges over the tabulated values of \cite{Arrington:2007ux} up to a given maximal value of $Q^2$, with $Q^2=0.1,0.2,\dots,1.2,1.4,1.6,1.8$ GeV$^2$. 

We chose our default bounds on the coefficients to be $|a_k| \leq 10,15$. The magnetic radius is then found by using eqn.(4). The extracted values are found to be independent of the number of coefficients we fit. Our default is 8 parameters. We present results in the Table 1 and 2 for two specific ranges of  $Q^2\leq 0.5$ GeV$^2$ and $Q^2\leq 1.0$ GeV$^2$.
\\
\begin{table}[h!]
\begin{minipage}{2.3 in}
\begin{tabular}{ | c | c | c | c |}
\hline
Bound on $a_k$ & $r_M^p$ & $+\sigma$ & $-\sigma$\\
\cline{1-4}
5 & 0.89 & 0.03 & 0.05\\
\cline{1-4}
10 & 0.91 & 0.03 & 0.06\\
\cline{1-4}
15 & 0.92 & 0.04 & 0.07\\
\cline{1-4}
20 & 0.93 & 0.04 & 0.07\\
\hline
\end{tabular}
\captionsetup{justification=centering}
\caption{$Q^2\leq 0.5$ GeV$^2$}
\end{minipage}\qquad
\hspace{2 cm}
\begin{minipage}{2.3 in}
\begin{tabular}{ | c | c | c | c |}
\hline
Bound on $a_k$ & $r_M^p$ & $+\sigma$ & $-\sigma$\\
\cline{1-4}
5 & 0.89 & 0.02 & 0.05\\
\cline{1-4}
10 & 0.91 & 0.03 & 0.07\\
\cline{1-4}
15 & 0.91 & 0.04 & 0.07\\
\cline{1-4}
20 & 0.91 & 0.05 & 0.08\\
\hline
\end{tabular}
\captionsetup{justification=centering}
\caption{$Q^2\leq 1.0$ GeV$^2$}

\end{minipage}
\end{table}

\subsection{Proton and neutron data}
Inclusion of the neutron data helps us to separate the isoscalar ($I=0$) and isovector ($I=1$) components of the proton form factor. For the neutron form factors ($G_M^n$) we use the data tabulated in \cite{Lung:1992bu,Gao:1994ud,Anklin:1994ae,Anklin:1998ae,Kubon:2001rj,Anderson:2006jp,Lachniet:2008qf}.

Again we observed that the central values of the extracted proton magnetic radius is consistent and do not depend on the number of coefficients we fit \cite{Epstein:2014zua} as shown in Table 3 and Table 4.

\begin{table}[h!]
\begin{minipage}{2.3 in}

\begin{tabular}{ | c | c | c | c |}
\hline
Bound on $a_k$ & $r_M^p$ & $+\sigma$ & $-\sigma$\\
\cline{1-4}
5 & 0.86 & 0.02 & 0.01\\
\cline{1-4}
10 & 0.87 & 0.04 & 0.05\\
\cline{1-4}
15 & 0.87 & 0.05 & 0.05\\
\cline{1-4}
20 & 0.88 & 0.04 & 0.06\\
\hline
\end{tabular}
\caption{$Q^2\leq 0.5$ GeV$^2$}
\end{minipage}\qquad
\hspace{2 cm}
\begin{minipage}{2.3 in}
\begin{tabular}{ | c | c | c | c |}
\hline
Bound on $a_k$ & $r_M^p$ & $+\sigma$ & $-\sigma$\\
\cline{1-4}
5 & 0.87 & 0.02 & 0.02\\
\cline{1-4}
10 & 0.88 & 0.02 & 0.05\\
\cline{1-4}
15 & 0.88 & 0.04 & 0.05\\
\cline{1-4}
20 & 0.88 & 0.05 & 0.06\\
\hline
\end{tabular}
\caption{$Q^2\leq 1.0$ GeV$^2$}
\end{minipage}

\end{table}

\subsection{Proton, neutron and $\pi\pi$ data}

Between the two-pion and four-pion threshold the only state that can contribute to the imaginary part of the magnetic isovector form factor is that of two pions. We can use the known information about ${\rm Im}\, G_M^{(1)}(t)$ in this region to raise the effective threshold of isovector form factor from $t_{\rm cut}=4m_\pi^2$ to  $t_{\rm cut}=16m_\pi^2$. The fitting is done by \cite{Hill:2010yb} 
\begin{equation}\label{Gcut}
G_M^{(1)}(t)=G_{\rm cut}(t)+\sum_k a^{(1)}_kz^k(t, t_{\rm cut}=16m_\pi^2,0). 
\end{equation}
$G_{\rm cut}(t)$ is calculated from eqn.(7).
Again, the extracted values are independent of the number of fitting parameters and are consistent over the range of $Q^2$. The error bars are also smaller. The results are presented in Tables 5 and 6.

\begin{table}[h!]
\begin{minipage}{2.2 in}
\begin{tabular}{ | c | c | c | c |}
\hline
Bound on $a_k$ & $r_M^p$ & $+\sigma$ & $-\sigma$\\
\cline{1-4}
5 & 0.867 & 0.010 & 0.013\\
\cline{1-4}
10 & 0.871 & 0.011 & 0.015\\
\cline{1-4}
15 & 0.873 & 0.012 & 0.016\\
\cline{1-4}
20 & 0.876 & 0.012 & 0.018\\
\hline
\end{tabular}
\caption{$Q^2\leq 0.5$ GeV$^2$}
\label{Table:1}
\end{minipage}\qquad
\hspace{2 cm}
\begin{minipage}{2.2 in}
\captionsetup{justification=centering}
\begin{tabular}{ | c | c | c | c |}
\hline
Bound on $a_k$ & $r_M^p$ & $+\sigma$ & $-\sigma$\\
\cline{1-4}
5 & 0.867 & 0.006 & 0.008\\
\cline{1-4}
10 & 0.874 & 0.008 & 0.015\\
\cline{1-4}
15 & 0.874 & 0.012 & 0.014\\
\cline{1-4}
20 & 0.875 & 0.013 & 0.016\\
\hline
\end{tabular}
\caption{$Q^2\leq 1.0$ GeV$^2$}
\end{minipage}

\end{table}

\section{Neutron magnetic radius extraction}

The magnetic radius of the neutron is defined as $r_M^n \equiv \sqrt{\langle r^2\rangle_M^n}$, where  

\begin{equation}
\langle r^2 \rangle_M^n=
\frac{6}{G_M^n(0)}\frac{d}{dq^2}G_M^n(q^2)\bigg|_{q^2=0}\,.
\end{equation}
We used the same technique and same data sets to extract the magnetic radius of neutron. Some results are shown in Table 7.

\begin{table}[h!]
\begin{center}
\begin{adjustbox}{width=\textwidth}
\begin{tabular}{ | m{1.6 cm} | c | m{1.5 cm} | m{1.2 cm} | c | m{1.5 cm} | m{1.2 cm} | c | m{1.5 cm} | m{1.2 cm} | c |}
\hline

$Q^2$(GeV$^2$) & Bound on $a_k$ & \multicolumn{3}{|c|}{Neutron data} & \multicolumn{3}{|c|}{Neutron and Proton data} & \multicolumn{3}{|c|}{Neutron,Proton and $\pi\pi$ data} \\
\hline
\cline{1-2}
& & $r_M^n$ & $+\sigma$ & $-\sigma$ & $r_M^n$ & $+\sigma$ & $-\sigma$ & $r_M^n$ & $+\sigma$& $-\sigma$\\
\cline{3-11}
\multirow{3}{.2 cm}{0.5}
 & 10 & 0.74 & 0.13 & 0.06 & 0.89&0.06 & 0.09 &0.89 &0.03 &0.03\\
\cline{2-11}
 & 15 & 0.65 & 0.21 & 0.07 &0.88 & 0.08 &0.09 &0.89 &0.03 &0.03 \\
\hline
1.0 & 10 & 0.77 & 0.17 & 0.09 &0.88 &0.06 &0.08 &0.88 &0.03 &0.01\\
\cline{2-11}
 & 15 & 0.74 & 0.20 & 0.11 &0.89 &0.07 &0.10 &0.88 &0.03 &0.02\\
\hline

\end{tabular}

\end{adjustbox}
\caption{Neutron magnetic radii for different data sets}
\end{center}

\end{table}

\section{Conclusion}

The purpose of this analysis was to try to resolve the discrepancies that exist in the literature regarding the extraction of the magnetic radius of the proton. To achieve that we used the \textquotedblleft $z$-expansion" method which incorporates the analytic structure of the form factors. 

Three data sets have been used for the extraction. From the proton data set we extracted the magnetic radius of the proton as $r_M^p = 0.91_{-0.06}^{+0.03}\pm 0.02$ fm. Inclusion of the neutron data gives the radius $r_M^p = 0.87_{-0.05}^{+0.04}\pm 0.01$ fm. When we add the $\pi\pi$ data along with these data sets the extracted value of the magnetic radius is $r_M^p = 0.87_{-0.02}^{+0.02}$ fm. Comparing our results with the PDG value $r_M^p =0.777\pm 0.014$ fm \cite{PDG:2014} we can say that our results are more consistent with the results of $r_M^p =0.854\pm 0.005$ fm extracted in \cite{Belushkin:2006qa} or $r_M^p =0.876\pm 0.020$ fm \cite{Borisyuk:2009mg} . Our study has also revealed that the extracted magnetic radius is independent of the number of parameters we fit or the range of $Q^2$ we used. We have reported all our results  with 8 parameters and two specific ranges $Q^2\leq 0.5$ GeV$^2$ and $Q^2\leq 1.0$ GeV$^2$.

The same procedure was applied to extract the magnetic radius of the neutron. Combining all three data sets (neutron, proton and $\pi\pi$) we extract the radius of the neutron to be $r_M^n =0.89\pm 0.03$ fm. Interestingly we notice that within the errors this value is consistent with the magnetic radius of the proton $r_M^p =0.87\pm 0.02$ fm.

\vskip 0.2in
\noindent
{\bf Acknowledgements}
\vskip 0.1in
\noindent
This work was supported by DOE grant DE-SC0007983.

\end{document}